\documentclass[twocolumn]{aastex62}

\usepackage{natbib, longtable}
\shorttitle{AA Tau}
\shortauthors{Covey, Larson, Herczeg \& Manara}

\begin{document}

\title{A Differential Measurement of Circumstellar Extinction for AA Tau's 2011 Dimming Event\footnote{Based on observations obtained with 
the Apache Point Observatory 3.5 m telescope, which is owned 
and operated by the Astrophysical Research Consortium.}}

\author[0000-0001-6914-7797]{K. R. Covey} 
\altaffiliation{The first two authors contributed equally to this work.}
\affiliation{Department of Physics \& Astronomy, Western Washington University, MS-9164, 516 High St., Bellingham, WA, 98225}

\author[0000-0002-9756-0383]{K. A. Larson} 
\altaffiliation{The first two authors contributed equally to this work.}
\affiliation{Department of Physics \& Astronomy, Western Washington University, MS-9164, 516 High St., Bellingham, WA, 98225}

\author[0000-0002-7154-6065]{G. J. Herczeg}
\affiliation{Kavli Institute for Astronomy and Astrophysics, Peking University, Yi He Yuan Lu 5, Haidian Qu, Beijing 100871, China}

\author[0000-0003-3562-262X]{C. F. Manara}
\affiliation{European Southern Observatory, Karl-Schwarzchild-Strasse 2, 85748 Garching bei M\"unchen, Germany}

%@arxiver{SED.eps,flux_ratio.eps,colormags_veiling.eps}

\begin{abstract}

\noindent AA Tau is a classical T Tauri star with a highly inclined, warped circumstellar disk.  For decades, AA Tau exhibited photometric and spectroscopic variability that were successfully modelled as occultations of the primary star by circumstellar material. In 2011, AA Tau entered an extended faint state, presumably due to enhanced levels of circumstellar dust. We use two sets of contemporaneous optical-NIR spectra of AA Tau, obtained in December of 2008 and 2014, to directly measure the wavelength-dependent extinction associated with the dust enhancement driving AA Tau's 2011 optical fade.  Taken alone, AA Tau's apparent optical-NIR increased extinction curve cannot be fit well with standard extinction laws. At optical wavelengths, AA Tau's dimming event is consistent with predictions of common extinction models for an increase of $A_V=2$, but no such model reproduces AA Tau's color-color excess at NIR wavelengths.  We show that veiling emission accounts for the apparent anomalous NIR extinction: after including this veiling flux, AA Tau's dimming event is consistent with a standard $A_V=2$ extinction law across the full optical-NIR range.  We also report an increase in AA Tau's mid-IR flux since its 2011 fade, and suggest that an increase in the height of AA Tau's inner disk is the most likely explanation for both the IR brightening and the additional extinction along the line of sight.  In addition to informing our understanding of AA Tau, this analysis  demonstrates that caution should be exercised when inferring extinction (and stellar parameters) from the NIR color-color excess of young stars with measurable NIR veiling fluxes.   

\end{abstract}


\section{Introduction}

Extinction corrections are a key component of studies of young stars, circumstellar 
disks, and star forming regions.  The ages and masses of protostars and T Tauri stars are typically inferred by comparing their temperatures and luminosities to the 
predictions of stellar evolutionary tracks. As \citet{Reggiani2011} show, 
uncertainties in the resulting age estimates are typically dominated by the 
uncertainty in the extinction correction applied to measure the stellar luminosity.  
Stellar accretion rates are most accurately measured from the luminosity of the 
Balmer continuum at $\lambda \lesssim 3700$ \AA\ \citep[e.g.][]{Gullbring1998}, where 
extinction's dramatic effect on the spectrum makes accurate reddening corrections 
essential to obtaining reliable results. Evaluation of the far-ultraviolet radiation of the disk, important for disk chemistry \citep{vanZadelhoff2003} and photoevaporation \citep[e.g.][]{Gorti2015}, is very sensitive to measurements of extinction and the extinction law \citep{McJunkin2014,McJunkin2016}.  
As these examples demonstrate, extinction 
corrections are a cornerstone for many studies that inform our understanding of the 
timescales and physical conditions of star and planet formation. 

Evidence has been building, however, that standard interstellar extinction laws may not provide reliable extinction corrections for the youngest stars. \citet{Herczeg2014} analyzed optical spectra of hundreds of T Tauri stars, determining extinctions by artificially reddening template spectra with a standard \citet{Fitzpatrick1999} extinction law to match the target star's continuum slope.  \citet{Herczeg2014} found good agreement with previous extinction estimates derived from optical spectra or colors \citep[e.g.][]{Kenyon1995}, but found their values were only about half as large as those inferred from near-infrared (NIR) spectra or colors \citep[e.g.][]{Fischer2011,Furlan2011}.  
A similar effect was seen by \citet{McJunkin2014}, whose UV-derived extinction estimates were yet lower than those derived from optical spectra.  
Some of these wavelength-dependent effects are likely introduced by the use of interstellar extinction curves for lines of sight that pass through the circumstellar disks.  Extinction curves may be leveraged to diagnose grain growth \citep[e.g.][]{Schneider2015,Guo2018} as a complement to sub-mm studies of colder regions of the disk \citep{Testi2014}.  However, wavelength-dependent differences could also be attributed to template colors that are mismatched with young stars \citep[e.g.][]{Gully2017} or to scattering, which dominates the optical and ultraviolet emission seen from stars with edge-on disks \citep[e.g.][]{White2004}.

The extinction to accreting young stars also often varies with time because of changes in disk structure or as disk warps rotate into our line of sight.  The archetype of the {\it dipper} class of extinction variables, AA Tau, had increased extinction as the accretion funnel flow and associated disk warp would rotate into our line of sight \citep{Bouvier2007}. These variations, however, typically occur on timescales of 2-10 days, comparable to the period of the star's rotation and the orbital timescale in the inner disk \citep[for more on the dipper class, see e.g., ][]{Cody2014}.  In late 2010, AA Tau entered a deep $\Delta V \sim 2-3$
photometric minimum \citep{Bouvier2013}, from which it has yet to fully emerge.  Because the \textit{direction} of the change in AA Tau's position in typical color-mag and color-color spaces agrees well with standard reddening vectors, the dimming event has been believed to be extinction-induced.  
The additional disk extinction
could be caused by an inflated inner disk rim at 0.1 AU or by a stable disk warp at several AU, favored by \citet{Bouvier2013} and \citet{Schneider2015}.
However, it is difficult to explain the system's full optical/NIR behavior with a standard interstellar extinction law: the \textit{magnitude} of the extinction changes inferred at NIR wavelengths are twice as large as those inferred from the optical. This disagreement between optical and NIR-based extinction inferences has been seen not just in AA Tau, but in the T Tauri studies referenced above as well, suggesting that we are witnessing a phenomenon that is common among T Tauri stars.   

The recent apparent increase in AA Tau's line-of-sight extinction  provides a unique opportunity to measure the detailed wavelength dependence of reddening due to circumstellar material. In this paper, we compare new optical-NIR spectra with spectra from AA Tau's earlier, less obscured state, similar to the optical-NIR analysis of RW Aur A by \citet{Koutoulaki2019}.  We measure the wavelength dependence of AA Tau's increased extinction in this period, and test if that change is consistent with predictions of standard interstellar extinction laws.
In addition to measuring the smooth continuum variations due to reddening, we also examine the absolute and relative strengths of emission and absorption features in AA Tau's spectrum, which limit any changes in accretion rate or stellar parameters (e.g. $T_{\rm eff}$) that may be affecting the spectrum as well. 
We use these features to show how veiling can explain the discrepancy between optical and NIR extinction.  We also discuss why inferring extinction in one band from color-color offset in other bands can be misleading and {\bf revisit} a less ambiguous technique using photospheric absorption lines, similar to that presented by \citet{Fischer2011}.

\section{Observations}

We obtained contemporaneous optical and NIR spectra of AA Tau in December 2008 and December 2014, before and well into the dimming event that began in late 2010.  We describe the acquisition and reduction of the optical and NIR spectra separately in sections \ref{sec:opt_spec} and \ref{sec:nir_spec}, respectively, and then discuss in section \ref{sec:merge_flux} how we combined the spectra obtained at each epoch into a contiguous, flux-calibrated spectrum spanning 0.3-2.5 $\mu$m. 

\subsection{Optical Spectra \label{sec:opt_spec} }

We analyze low- and high-resolution optical spectra of AA Tau collected by multi-year monitoring programs that cover AA Tau's 2011 fade. The low-resolution spectra were obtained in December 2008 with the Double Spectrograph \citep[DBSP; ][]{Oke1982} on the Hale 200 inch telescope at Palomar Observatory, and in December 2014 with the SuperNova Integral Field Spectrograph \citep[SNIFS, ][]{Aldering2002, Lantz2004} on the University of Hawaii 2.2m telescope on Mauna Kea. We also analyze high-resolution spectra obtained with the EsPaDoNs spectrograph between December 2008 and January 2015. We discuss each of these three datasets in turn below.

\subsubsection{2008 Palomar Observations} 

AA Tau was observed at 2:00 UT on 2008 December 28 as a part of the comprehensive survey of T Tauri stars whose first results were reported by \citet{Herczeg2014}.  For the full details of the configuration of the DBSP spectrograph, and the acquisition and reduction of these spectra, we direct the reader to the complete description provided by \citet{Herczeg2014}. In brief, the spectra were observed with a 4\arcsec\ slit and the 2048 $\times$ 4096 CCD 23 and 1024 $\times$ 1024 Tektronix detectors on the spectrograph's blue and red arms respectively. Red spectra were taken as a series of 2s exposures; blue spectra were recorded simultaneously with a smaller number of 60s exposures.  Exposures were overscan subtracted, flat-fielded, and extracted using a 21 pixel window centered on the source's trace while subtracting sky emission measured in a nearby set of columns. The spectrum was then corrected for light losses based on the measured seeing as a function of wavelength, assuming a Gaussian point spread function. 

\subsubsection{2014 SNIFS Observations}

Using 225 microlenses that each sample an area of 0.4$\times$0.4\arcsec\ region of the sky, and a dichoric to split each beams' light into separate blue (3200-5600 \AA, R$\sim$1000) and red (5200-10000 \AA, R$\sim$1300) spectrographs, SNIFS produces combined R$\sim$1000 spectra that cover wavelengths of 3200-9700 \AA\ and fully populate a 6$\times$6\arcsec~footprint on the sky. AA Tau was observed at 07:45 and 12:45 (UT) on 2014 December 2 (obs. spec-C14-336-61 and spec-C14-336-162 respectively) and on 2014 December 12 at 06:34 and 13:17 (UT; 346-66 and 346-199 respectively) with 200s integration times.  Spectra were reduced with the automated SuperNova Factory spectral reduction pipeline \citep{Bacon2001, Aldering2006}, which performs dark, bias, flat-field, bad pixel and cosmic ray correction before extracting and wavelength calibrating the spectrum using arc lamps taken close in time to the science observations. Flux calibration was performed using archival sensitivity measurements, and adjusted based on contemporaneous photometric measurements.

\subsubsection{2008-2015 ESPaDOnS Observations}

To enable accurate measurements of AA Tau's red-optical veiling, 227 observations of AA Tau collected between December 2008 and January 2015 with the Echelle SpectroPolarimetric Device for the Observation of Stars \citep[ESPaDOnS; ][]{Donati2003,Silvester2012} were retrieved from the CADC interface to the archive of the 3.6m Canada-France-Hawaii Telescope (CFHT). These observations were made using ESPaDOnS spectropolarimetric mode, but {\bf we} consider here only the total intensity spectra, which provide coverage from 3700-10290~\AA\ across 40 spectral orders with a typical spectral resolution of 68,000. Archive observations are reduced with the Libre-ESpRIT package \citep{Donati1997}, which reports typical signal-to-noise ratios per resolution element of about 55 at 6660 \AA\ in late 2008 and early 2009 from a single 1200s exposure; after AA Tau faded in 2011, typical signal-to-noise ratios declined to about 30 at 6660 \AA\ in a somewhat longer 1500s exposure.  The set of observations from December 2008 and January 2009 were published by \citet{Donati2010}.

\subsection{Near-Infrared Spectra \label{sec:nir_spec}}

NIR spectra of AA Tau were acquired in 2008 and 2014 with the TripleSpec spectrograph on the Astrophysical Research Consortium (ARC) 3.5 meter telescope at Apache Point Observatory in Sunspot, New Mexico.  The TripleSpec NIR spectrograph provides simultaneous coverage of nearly the full wavelength range from 0.95 $\mu$m to 2.46 $\mu$m \citep{Wilson2004}.  Observations were made at 01:30 UT on 2008 December 29, and three times each night on 2014 December 2 and 12.  Spectra were obtained with a 1.1$\arcsec$ slit to achieve a resolution of R $\sim$ 3200, and using an ABBA dither sequence to allow the removal of sky emission by differencing sequential exposures.  Integration times of 60s or 120s per dither position were adopted  depending on the seeing, which determined the fraction of the star's light which would enter the 1.1$\arcsec$ slit, and thus the S/N that would result from a given integration time. 

All spectra were reduced using the IDL-based SpeXTool pipeline, originally developed by \citet{Cushing2004}, and modified for use with TripleSpec data.  Spectra were differenced, flattened, extracted, and wavelength calibrated prior to telluric correction and flux calibration; these corrections were performed using a spectrum of a nearby A0V star obtained at similar air mass and time to each target, and derived using the \texttt{xtellcorr} IDL package \citep{Vacca2003}. 

\begin{figure}[tb] 
    \centerline{\includegraphics[angle=90,width=1.1\columnwidth]{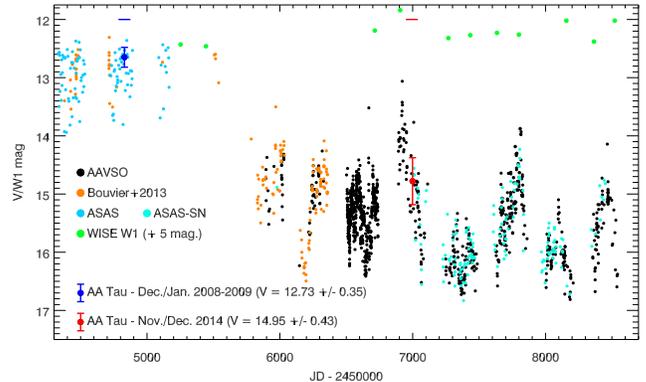}}
    \caption{V band photometry of AA Tau obtained between mid-2007 and late 2015. Data shown were collected by members of the American Association of Variable Star Observers (AAVSO; black circles), \citet{Bouvier2013} (orange circles), the All-Sky Automated Survey \citep[ASAS; blue circles; ][]{Pojmanski1997}. } 
    \label{fig:AATau_LC}
\end{figure}

\subsection{Optical \& Mid-IR photometry \label{sec:midir}}

To characterize AA Tau's photometric variations over the course of its 2011 dimming event, we have assembled archival optical and mid-IR photometry of AA Tau, as shown in Figure  \ref{fig:AATau_LC}.  The optical photometry consists of calibrated photometric measurements collected by members of the American Association of Variable Star Observers (AAVSO), \citet{Bouvier2013}, the All-Sky Automated Survey \citep[ASAS, ][]{Pojmanski1997}, and the All-Sky Automated Survey for SuperNovae \citep[ASAS-SN, ][]{Shappee2014, Kochanek2017}. 

The WISE \citep{wright2010} and NEOWISE \citep{mainzer2011} missions have obtained photometry in $W1$ ($\sim 3.6$ $\mu$m) and $W2$ ($\sim 4.5$ $\mu$m) bandpasses about every 6 months.  Each visit includes between 10--15 separate integrations.  We obtained WISE and NEOWISE photometry \citep{cutri2014} from the NASA/IPAC Infrared Science Archive.  Table~\ref{tab:wise} lists the WISE photometry from each epoch, obtained by calculating the median of each individual integration.  Table~\ref{tab:wise} also includes the {\it Spitzer} IRAC1 and IRAC2 photometry of AA Tau \citep{rebull2010}, which cover similar bands as $W1$ and $W2$.  

To enable a direct comparison of these optical and mid-IR photometry, we compute AA Tau's approximate $V$-band magnitude at the time of each mid-IR epoch.  For all epochs in Table \ref{tab:wise} with MJD $>$ 56000, we report the median of all AAVSO $V$-band measurements collected within 10 days of the mid-IR photometry.  For the first three mid-IR epochs (MJDs $<$ 56000), where light curves are more sparsely populated, we report the median magnitude of all $V$-band photometry collected by ASAS \& \citet{Bouvier2013} within 150 days of the mid-IR epoch.

\begin{table}[]
\caption{Mid-IR photometry}
\label{tab:wise}
 \begin{tabular}{cccccc}
 \hline
    MJD    &   W1   & Err  & W2  & Err & $<V>$\\
    \hline
 53421.5  &   (7.29) & 0.05 &  (6.84) & 0.05 & 12.89 \\
 55252.5  &   7.42  &  0.02   &    6.76  &0.02 & 12.66\\
 55443.6  &   7.49  &  0.04   &    6.77  &0.02 & 12.68\\
 56714.6  &   7.14  &  0.02   &    6.40  &0.01 & 15.14\\
 56905.6  &   6.82  &  0.02   &    6.23  &0.01 & 14.18 \\
 57073.6  &   6.71  &  0.03   &    5.88  &0.02 & 16.17 \\
 57268.4  &   7.31  &  0.02   &    6.45  &0.01 & 16.53 \\
 57434.1  &   7.25  &  0.02   &    6.41  &0.01 & 15.96 \\
 57634.6  &   7.24  &  0.02   &    6.33  &0.01 & 16.06 \\
 57799.9  &   7.20  &  0.02   &    6.45  &0.01 & 14.71 \\
 57998.8  &   6.64  &  0.03   &    5.88  &0.02 & 16.48 \\
 58156.9  &   7.00  &  0.02   &    6.26  &0.01 & 16.23 \\
 58364.4  &   7.36  &  0.02   &    6.51  &0.01 & 16.10 \\
 58521.3  &   7.02  &  0.02   &    6.26  &0.01 & 15.66 \\
 %58730.0  &   7.16  &  0.02   &    6.41  &0.01 & \\
 \hline
 \multicolumn{6}{l}{MJD and Errors are medians of each observation}\\
 \multicolumn{6}{l}{~~~within a given epoch}\\
 \end{tabular}
 \end{table}

\section{Analysis}

\subsection{Producing Merged Optical/NIR Spectra \label{sec:merge_flux} }

To create contiguous spectra, we first scale the individual optical spectra to a common continuum and average the spectra for each night.  For December 2, no scaling is necessary in the blue spectra and one red spectrum is scaled up by 5\% to match the continua from 0.78 $\mu$m to 0.92 $\mu$m of the others.  For December 12, one of the two spectra in the blue is much noisier than the other, and is therefore not included in further analysis. In the red band, one spectrum is scaled up by 15\% to match the other.  Overlap between the blue and red band is negligible, so no scaling is attempted between the two bands, but we note that the two bands appear to be better matched in the December 2 data than they do in the December 12 data. Spectra are shifted very slightly in wavelength to line up prominent emission features, less than 0.005 $\mu$m. Resulting scaled and shifted spectra are then averaged, resulting in one optical spectrum for each night.  

We perform similar scaling in the infrared.  Two of the three infrared spectra from 2014 December 2 agree well with each other.  A simple scaling of 5\% matches the continua from 0.98 $\mu$m to 1.28 $\mu$m.   We discard a third spectrum taken at higher airmass (1.33 vs.\ 1.1) and under non-photometric conditions, as it has a a steeper continuum slope (i.e. 20\% redder at 2.5 $\mu$m) than the two earlier spectra.  For the December 12 data, two of the infrared spectra are scaled by 26\% and 12\% to match a third spectrum in the same continuum range.  Scaled infrared spectra are then averaged, resulting in one infrared spectrum for each night. 

The 2008 infrared and optical spectra do not overlap, but they appear to match well, so no additional scaling of the 2008 infrared spectrum is necessary.  To match the 2014 infrared with corresponding optical spectra, we fit a line to the long-wavelength optical continuum between 0.85 $\mu$m to 0.90 $\mu$m and project that line to the short-wavelength infrared at 0.945 $\mu$m.  The ratio of that value to the mean flux detected in the infrared spectrum between 0.94 $\mu$m and 0.95 $\mu$m provides a scaling factor that we apply to the optical spectrum to bring the two into agreement.  To match the optical to the infrared, the optical spectrum for December 2 is reduced by 0.7\%.  A scale factor so close to unity adds confidence to the flux calibration of the observational procedure.  

\begin{figure}[!htb] 
    \centerline{\includegraphics[angle=0,width=\columnwidth]{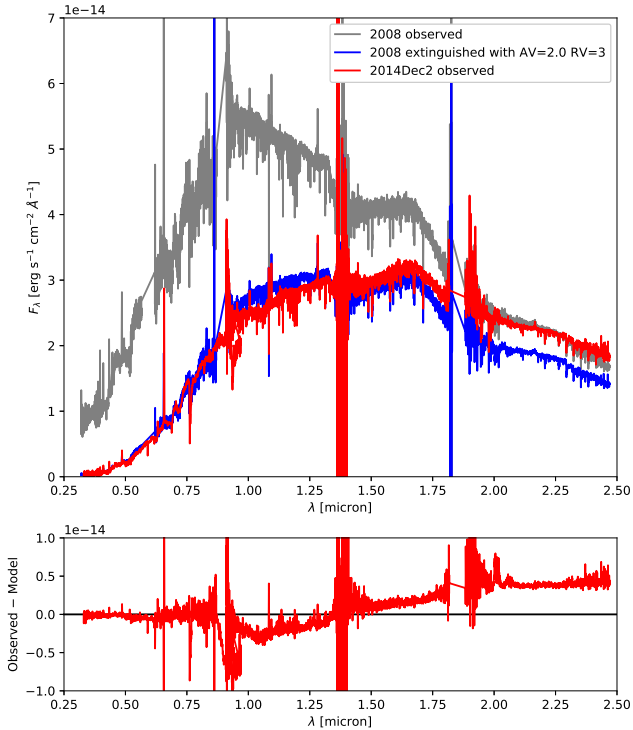}}
    \caption{Spectra of AA Tau.  Flux in 2008 before the dimming event is shown in grey, on December 2 in 2014 after the dimming event in red.  For comparison, the 2008 data extinguished by a $A_V=2$ and $R_V=3$ model of \citet{Fitzpatrick1999} is shown in blue.  The difference between the model and the 2014 observations in the bottom panel shows a systematically different shape between the two curves in the infrared.} 
    \label{fig:AATau_spectra}
\end{figure}

\begin{figure}[!htb] 
    \centerline{\includegraphics[angle=0,width=\columnwidth]{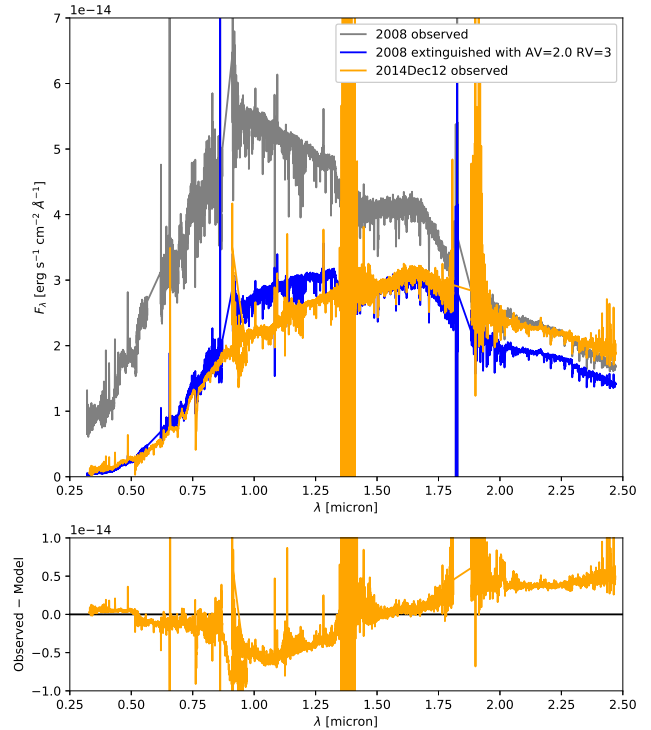}}
    \caption{Same as Figure \ref{fig:AATau_spectra}, but for 2014 December 12 in orange.} 
    \label{fig:AATau_spectra2}
\end{figure}

The final flux spectra of AA Tau in 2008 (grey) are shown in comparison to 2014 December 2 (red) and 2014 December 12 (orange) in Figure \ref{fig:AATau_spectra} and Figure \ref{fig:AATau_spectra2}, respectively.  \citet{Bouvier2013} report that AA Tau dimmed $A_V \sim 2.0$ from 2008 to 2014, which we confirm with both archive photometry and synthetic photometry of the spectra in the next section.  Therefore, in Figure \ref{fig:AATau_spectra} and Figure \ref{fig:AATau_spectra2}, we also include the 2008 spectrum extinguished by $A_V=2$ (blue).  We use the extinction model of \citet[][hereafter F99]{Fitzpatrick1999}, which parameterizes interstellar extinction with visual extinction, $A_V$, and the ratio of extinction to blue-visual color excess, $R_V=A_V/E(B-V)$.  The parameter $R_V$ measures the shape of the blue-visual normalized extinction curve, with $R_V=3$ typical of the diffuse interstellar medium.  The bottom panels of Figure \ref{fig:AATau_spectra} and Figure \ref{fig:AATau_spectra2} show the difference between the 2014 observed spectra and the extinguished 2008 spectrum.  In both cases, the extinguished 2008 spectrum is clearly not an acceptable model for the 2014 spectrum, especially in the near infrared.

\subsection{Synthetic Photometry from Spectra}

We perform synthetic photometry of the 2008 and two 2014 spectra with \texttt{synphot} \citep{synphot}, an affiliated package of Astropy; resulting magnitudes using Johnson $BRVI$ and Bessel $JHK$ filter curves are listed in Table \ref{tab:AATau_Phot}.  To test flux calibration of the 2008 spectrum, we compare results of synthetic photometry with the archival photometric data introduced in \ref{sec:midir}.   From the lightcurve, we estimate $V = 12.73 \pm 0.35$ for AA Tau in December/January 2008/2009, which agrees with our synthetic photometry result of $V = 12.85$ and confirms the flux calibration of the optical 2008 spectrum. Similarly, we estimate of $V = 14.95 \pm 0.43$ from the lightcurve in Figure \ref{fig:AATau_LC} for November/December 2014, which agrees with our synthetic photometry of $V = 14.85$ and 14.95 for the two nights in late 2014.  

\begin{deluxetable}{rrrrrrrr}
\tablewidth{\columnwidth}
\tabletypesize{\scriptsize}
\tablecaption{Synthetic Photometry from AA Tau Spectra\label{tab:AATau_Phot}}
\tablehead{ 
\colhead{Date} & \colhead{$B$} & \colhead{$V$} & \colhead{$R$} & \colhead{$I$} & \colhead{$J$} & \colhead{$H$} & \colhead{$K$} 
}
\startdata
2008Dec28 & 13.99 & 12.85 & 11.57 & 10.54 &  9.48 &  8.64 &  8.13 \\
2014Dec02 & 16.59 & 14.85 & 12.84 & 11.50 & 10.04 &  8.90 &  8.12 \\
2014Dec12 & 16.18 & 14.95 & 12.95 & 11.61 & 10.07 &  8.92 &  8.11 
\enddata
\end{deluxetable}

\begin{figure}[!htb] 
    \centerline{\includegraphics[angle=0,width=\columnwidth]{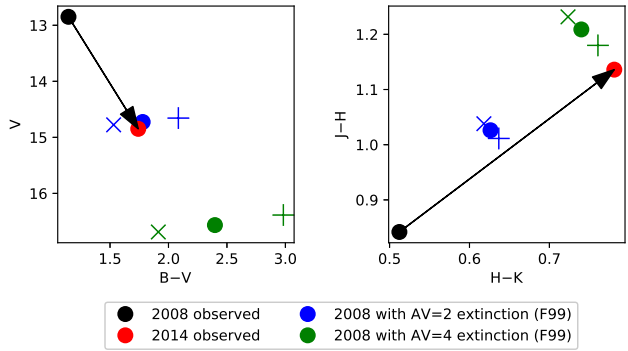}}
    \caption{Synthetic photometry from the AA Tau spectra shown as optical color-magnitude and NIR color-color diagrams.  The black and red dots are photometry calculated from the 2008 and 2014 December 2 spectra, respectively, The black vector represents the empirical reddening vector for the dimming event.  Synthetic photometry from the 2008 spectrum reddened by the \citet{Fitzpatrick1999} extinction curves for $A_V=2$ (blue) and $A_V=4$ (green) are shown for comparison, with $R_V=2$, 3, and 5 as plus, circle, and cross symbols respectively.} 
    \label{fig:AATau_optnirplot}
\end{figure}

Both synthetic photometry of the spectra and the archive data light curve imply that AA Tau dimmed by $\Delta V=2.0$ between the two epochs captured by our combined optical-NIR spectra. This $\Delta V=2.0$ mag variation is notably larger than the $\Delta V \sim$ 1 mag variations that are visible in AA Tau's has exhibited on shorter timescales (days to weeks), both before and after its late 2010 fade.  These short timescale variations, visible in the historical light curve shown in Figure \ref{fig:AATau_LC} and elsewhere in the literature, are likely captured at unknown phases in our spectra, but the $\Delta V \sim$ 2 mag difference in the synthetic photometry suggests that AA Tau's long-lived dimming event drives most of the differences observed between the two epochs sampled by our spectra.  This is also supported by the presence of a clear separation at V$\sim$14 mag between the bright and faint states; our optical-NIR spectral epochs lie $\gtrsim 1$ mag away from this dividing line, suggesting that our spectral epochs capture the dominant behavior in each state.

To investigate what these spectra reveal about AA Tau's photometric behavior during this period, we show an optical color-magnitude plot with synthetic photometry in the left panel of Figure \ref{fig:AATau_optnirplot}.  The arrow drawn from 2008 to 2014 December 2 photometry is the empirical reddening vector between these two observations.  For comparison, we also plot the effect of applying extinction models to the 2008 spectra and repeating synthetic photometry.  The F99 model shown here parameterizes extinction curves with $A_V$ and $R_V = A_V/E(B-V)$, which represent the amount of visual extinction and the shape of the normalized blue-visual extinction curve, respectively.  Blue points in Figure \ref{fig:AATau_optnirplot} show the effect of extinguishing the 2008 spectrum with an $A_V=2$ extinction model; green shows the effect of $A_V=4$.  Different symbol shapes represent different values of the parameter $R_V$ in the extinction model. The circles are the $R_V=3$ curve shape, typical of the average diffuse interstellar medium.  The cross symbols are the $R_V=5$ curve shape, representing large-$R_V$ extinction that is found in dense clouds with grain growth due to coagulation.  The plus symbols are the $R_V=2$ curve shape, which implies an enhanced relative abundance of small grains.  Parameterizing the extinction curves with $R_V$ reflects the empirical evidence that variations in normalized extinction curves tend to be correlated across wavelength. 

The optical color-magnitude plot in the left panel shows that AA Tau's dimming event is consistent with $A_V=2$ and the standard value of $R_V = 3$.  The empirical reddening vector in the NIR color-color plot in the right panel, on the other hand, appears to be consistent with synthetic photometry calculated from models parameterized by $A_V=4$ and outside the range of expected values of $R_V$.   This observation of mismatch between the optical and NIR reddening vector before and after the dimming event, similar to the observations of  \citet[][ Figure 3]{Bouvier2013} and \citet[][ Figure 5]{Schneider2015}, is the central motivation for this project.

\subsection{Spectral Absorption Lines\label{sec:lines}}

This section describes measurements of the equivalent width (EW) of photospheric absorption lines.  We use custom IDL routines to measure the absolute absorbed flux ($I$) of the lines using trapezoidal integration between two continuum points on the red and blue end of each line. We calculate the EW of each line by dividing the absolute integrated absorbed flux,
\begin{equation}
    I = \int_{\rm line} \left( F_{\rm cont}- F_\lambda \right) d\lambda
\end{equation}
by the average flux in the two continuum regions on either side of the line ($F_{\rm cont}$), such that
\begin{equation}
EW = \frac{I}{F_{\rm cont}}.  \label{eq:EWdef}
\end{equation}
A featureless veiling flux ($E$) elevates the continuum that an absorption line is measured relative to, and thus decreases the strength of that line when measured as an equivalent width, 
\begin{equation}
EW_{\rm veiled} = \frac{I}{F_{\rm cont}} = \frac{I}{F_{*} + E} < \frac{I}{F_{*}} = EW_{\rm unveiled}
\end{equation}

To quantify the contribution of any non-photospheric veiling flux in AA Tau's NIR spectrum, we manually measured the strengths of 23 photospheric absorption lines identified by \citet{Covey2010} as useful temperature and log $g$ indicators in IRTF/SpeX observations of young stellar objects. These lines span 1 $\mu$m to 2.3 $\mu$m, the full range sampled by IRTF/SpeX and TripleSpec,  with somewhat denser sampling at shorter wavelengths.  Repeating this process multiple times provided empirical estimates of the error in each line measurement due to systematic differences in the selected continuum regions. In addition to the TripleSpec data that are our primary focus, we also performed veiling measurements on an archival IRTF/SpeX observation of AA Tau obtained by \citet{Fischer2011} in 2006, and a re-reduced X-Shooter observation obtained in 2012 by \citet{Bouvier2013} and degraded to match the resolution of the TripleSpec observations.  

At low-resolution, AA Tau's optical spectrum is dominated by broad molecular absorption bands that do not lend themselves to veiling measurements. High-resolution optical spectra provide access to narrow, well-resolved optical line strength measurements, from which more reliable optical veiling measurements can be obtained.  We therefore used a modified version of the IDL routine described in the previous section to measure absolute line fluxes and EWs of 10 strong photospheric absorption lines, ranging from the 5270 \AA\ MgH band to the 8689 \AA\ calcium triplet, which are present in ESPaDOnS spectra of AA Tau.  Relying on the consistent wavelength calibration of the ESPaDOnS data, we automated the routine to compute absolute line fluxes and EWs for a 1-3 \AA\ region including each line, using custom selected 1-3 \AA\ continuum regions on either side of, and not necessarily immediately adjacent to, each absorption line.  The exceptions are the 5660 \AA\ and 5270 \AA\ features, which are moderately wide (17-64 \AA\ ) absorption bands which nonetheless provide useful constraints on the optical flux at the bluest wavelengths. 
Line measurements are presented in Table \ref{tab:Opt_Veil}. 

\subsection{Accretion-Sensitive Emission Lines}

\begin{figure}[!htb] 
\centerline{\includegraphics[angle=0,width=\columnwidth]{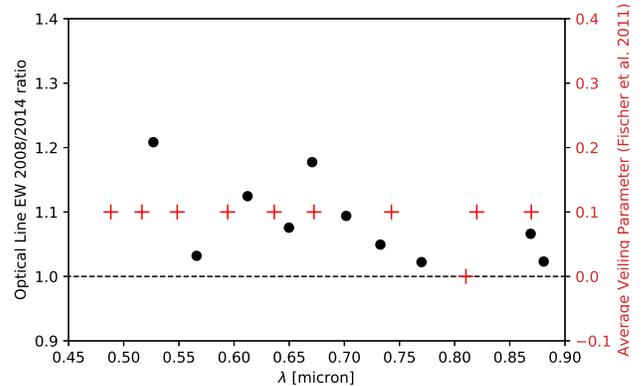}}
    \caption{Equivalent width ratio of photospheric absorption lines measured from ESPaDoNS spectra obtained before and after 2011 (black dots).  Also shown for comparison as red dots, with an explicit conversion to veiling flux on the right hand axis, is the average 10\% veiling flux measured in the optical {\bf by \citet{Fischer2011}.}} 
\label{fig:opt_veiling}
\end{figure}

To quantify potential changes in AA Tau's accretion rate, we measure the EWs and absolute integrated fluxes of the accretion sensitive H$\alpha$ and H$\beta$ emission features in each of the spectra we analyze. The absolute fluxes, presented in Table \ref{tab:AATau_EqW}, indicate that AA Tau's H$\alpha$ and H$\beta$ emission lines declined by a factor of $\sim$4-6 as the star's overall optical brightness faded from 2008 into 2014. The local continuum appears to have declined more significantly than the line fluxes, however, such that the lines' EWs increased by a factor of $\sim$2 over the same period. This increase in EW could be explained by either the addition of differential extinction between the stellar photosphere and emission line region, which surpresses the local continuum more effectively than the emission flux, or by a comparable rise in AA Tau's accretion rate between 2008 and 2014.  A near doubling of the accretion rate would produce measurable changes in AA Tau's emission lines, but would be unlikely to drive measurable changes in the star's broader continuum at the red-optical to NIR wavelengths that are the primary focus of our work.  More dramatic changes would be present in the continuum at blue-optical and shorter wavelengths, but also in line or bound-free emission that traces AA Tau's accretion activity such as its Balmer jump, HeI 5876 line emission, or optical veiling estimates inferred from the the EW ratio of photospheric lines in ESPaDoNS data obtained before and after 2011, whose measurements are described in the previous section and shown in Fig. \ref{fig:opt_veiling}.  In the analysis that follows, we assume that AA Tau's accretion rate increased by no more than a factor of two between 2008 and 2014, but note that measurements at bluer wavelengths would provide better constraints on potential changes in AA Tau's accretion rate.\\

\begin{deluxetable}{rrrrr}
\tablewidth{\columnwidth}
\tabletypesize{\scriptsize}
\tablecaption{Equivalent Widths of AA Tau's Accretion Sensitive Emission Lines\label{tab:AATau_EqW}}
\tablehead{ 
\colhead{Obs.} & \colhead{H$\beta$} & \colhead{H$\beta$} & \colhead{H$\alpha$} & \colhead{H$\alpha$} \\
\colhead{Date} & \colhead{EqW} & \colhead{Flux} & \colhead{EqW} & \colhead{Flux}
}
\startdata
12-28-2008 & 5.45 \AA & 9.95E$-$18 & 20.4 \AA & 6.8E$-$17 \\
2-25-2012 & 11.02 \AA & 2.1E$-$18 & 32.1 \AA & 1.2E$-$17\\
12-02-2014 & 9.04 \AA & 1.9E$-$18 & 30.0 \AA & 2.5E$-$17\\
12-12-2014 & 12.9 \AA & 3.4E$-$18 & 49.5 \AA & 3.6E$-$17 \\
\enddata
\end{deluxetable}

\begin{figure*}[!tb] 
    \centerline{\includegraphics[angle=0,width=2\columnwidth]{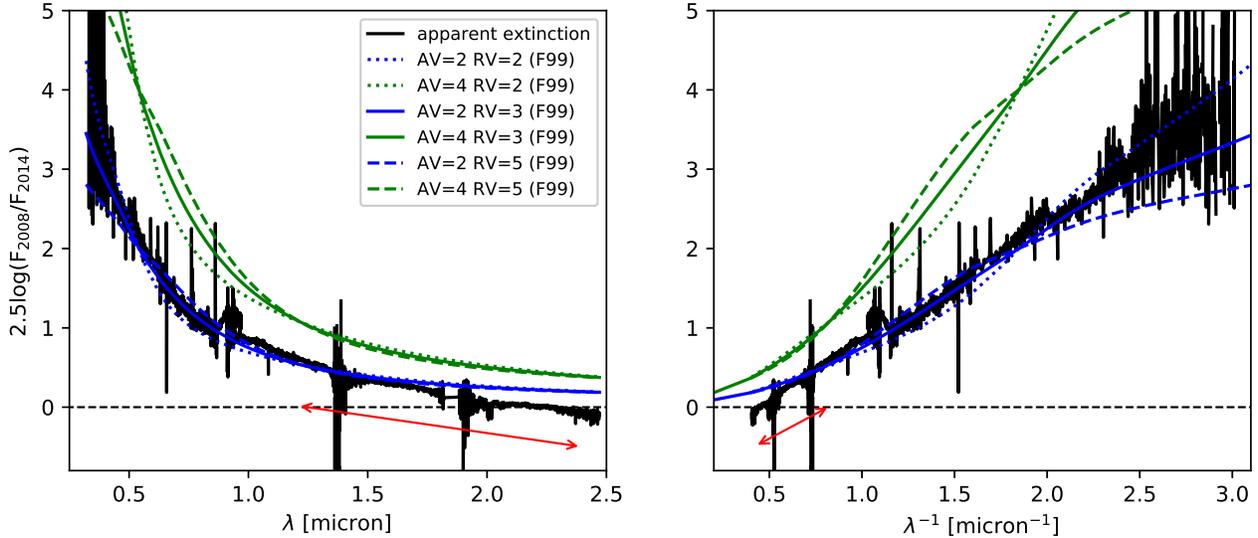}}
    \caption{Spectral flux ratio of AA Tau's dimming event.  In log space, the flux ratio is proportional to apparent extinction.  \citet{Fitzpatrick1999} $A_V=2$ (blue) and $A_V=4$ (green) models from are shown for comparison.  Solid lines are $R_V=3$ models, dotted lines are $R_V=2$ models, and dashed lines are $R_V=5$ models. The right plot is the same as the left, except plotted versus inverse wavelength.  Note that the green curves better fit the average slope of the NIR flux ratio, emphasized by the red arrows.  } 
    \label{fig:AATau_ext}
\end{figure*}

\section{Analysis}

\subsection{Neither $A_V=2$ nor $A_V=4$ reproduces AA Tau's 2008-2014 differential extinction spectrum}

Our goal in this analysis is to resolve the  contradiction between the apparent optical and NIR extinction increases of Figure \ref{fig:AATau_optnirplot}.  Extinction of starlight is caused by scattering and absorption by dust grains in the line of sight.  Extinction is usually expressed as a difference in magnitudes between an object with dust and an object without dust.  We can calculate a wavelength-dependent extinction curve of AA Tau's dimming event with the spectra of AA Tau before and after the event, such that 
\begin{equation}
    A_\lambda = -2.5\log\left( \frac{F_{\lambda, 2014}}{F_{\lambda, 2008}} \right). \label{eq:ext_curve}
\end{equation}
Magnitude extinction is not calculated from the $A_\lambda$ curve, but is obtained by integrating flux over a filter bandpass in the usual way.

We show the empirical differential extinction curve for AA Tau between the two epochs of our observations in Figure \ref{fig:AATau_ext}, with the six extinction curves used in Figure \ref{fig:AATau_optnirplot} in the same color scheme (blue for $A_V=2$ and green for $A_V=4$, and linestyles denoting distinct values of $R_V$). The left and right panels of Figure \ref{fig:AATau_ext} show the same spectral data plotted versus wavelength and inverse wavelength, respectively.  Plotting versus inverse wavelength or wavenumber is common practice in the literature of the interstellar medium research community because it emphasizes that extinction models typically have a y-intercept of zero, that is, approach zero extinction at very long wavelengths. Notably, none of the blue $A_V=2$ or green $A_V=4$ extinction models over a range of $R_V$ values can reproduce AA Tau's empirically inferred extinction spectrum.  The failure of the $A_V=4$ models is clear, with more extinction than is observed at all wavelengths.   The $A_V=2$ models' failure is less prominent, but emerges at wavelengths longer than 2 $\mu$m, where the model  overpredicts the extinction by a smaller, but nonetheless significant, amount.

\subsection{NIR color excesses diagnose the slope, but not amount, of the NIR extinction}

Before discussing the reason that none of the model curves in Figure \ref{fig:AATau_ext} are a good fit to the empirical AA Tau extinction between these two epochs, we pause to explain why the AA Tau dimming event appears consistent with an increase of $A_V=4$ in the NIR color-color plot (right panel) of Figure \ref{fig:AATau_optnirplot}, given that the $A_V=4$ curves (green) seem almost unrelated to the empirical AA Tau data in Figure \ref{fig:AATau_ext}.  
The key is understanding how color excess is related to extinction.  Color excess (or reddening) is calculated as the difference between a color observed with dust in the sightline and the color inferred without dust in the sightline.  Color excess is equivalent to differential extinction, 
\begin{equation}
    E(\lambda-{\rm ref}) = A_\lambda-A_{\rm ref}.
\end{equation}  
In a color-color plot, the vertical and horizontal offsets between an intrinsic and an observed point are color excess.\footnote{Unfortunately, $E$ is used for both veiling excess emission and color excess due to dust.  In this paper, $E$ with a single subscript is excess emission, while $E$ as a function of a difference is reddening.}  

Large color excess or reddening means a large difference between extinction in those bands.  In other words, large color excess means a steep extinction curve. A careful examination of Figure \ref{fig:AATau_ext}  shows that the \textit{steepness} of AA Tau's extinction in the near infrared (emphasized by red arrows) is better represented by the model parameterized by $A_V=4$ (green) than the model with $A_V=2$ (blue). Put more simply: in Figure \ref{fig:AATau_ext}, the blue $A_V=2$ extinction models are noticeably flatter than the empirical measurements at NIR wavelengths but the the green $A_V=4$ model has a similar slope.  The large displacement in color-color space in Figure \ref{fig:AATau_optnirplot} is caused by the steep apparent NIR extinction curve, not a large amount of extinction overall, so is best reproduced by the steeper $A_V=4$ extinction models, rather than the flatter $A_V=2$ models.

\begin{figure}[!tb] 
    \centerline{\includegraphics[angle=0,width=\columnwidth]{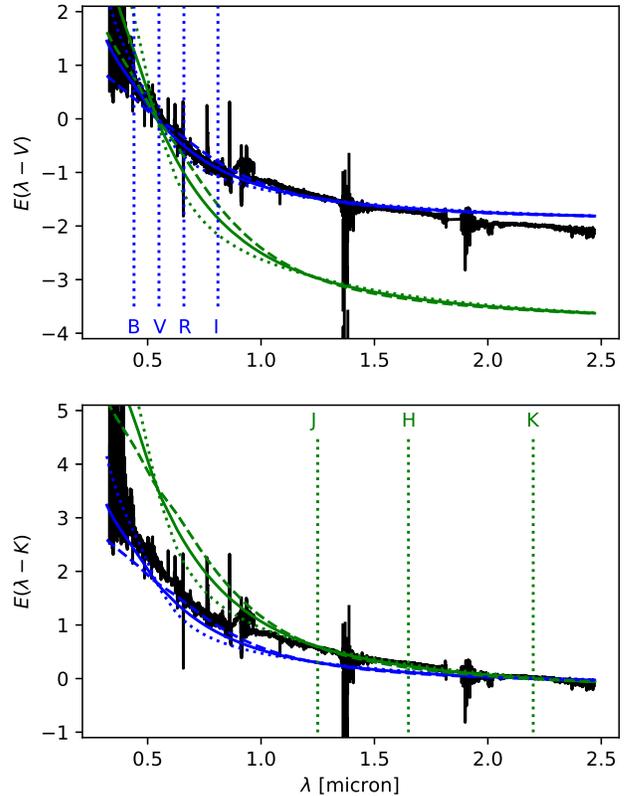}}
    \caption{Apparent reddening of AA Tau relative to $V$ and $K$. Green and blue curves have same meaning as in Figure \ref{fig:AATau_ext}.  In log space, the flux ratio relative to a reference passband is apparent color excess.  Notice that the $A_V=2$ curves (blue) fit color excess relative to $V$ in optical passbands (blue vertical lines), while the $A_V=4$ curves (green) are a better approximation to the color excess relative to $K$ in the NIR passbands (green vertical lines).}
    \label{fig:AATau_reddening}
\end{figure}

Inferring extinction from color excess can be difficult, especially in passbands other than those where color excess is calculated.  To further illustrate why, we plot the apparent extinction from Figure \ref{fig:AATau_ext} as apparent color excess relative to both $V$ and $K$ in the top and bottom panels of Figure \ref{fig:AATau_reddening}, respectively.  Color excess curves are offset vertically to pass through zero at the reference passband and have a long-wavelength limit equal to the negative of the extinction in the reference passband.  Careful investigation of both panels in Figure \ref{fig:AATau_reddening} shows that the shape of the $A_V=4$ models (green) are qualitatively similar to AA Tau in the NIR passbands (green vertical lines) simply because the reddening curve in the near infrared is steep, not because there is actually four magnitudes of extinction in the optical passbands.  Color excess relative to $K$ has a shape consistent with the steeper models.  Interestingly, note that for the $J$ band, color excess relative to $V$ is consistent with $A_V=2$ models, while color excess relative to $K$ is consistent with the $A_V=4$ models.  Also, because NIR extinction is so much less than optical extinction, the quantitative difference between the $A_V=2$ and $A_V=4$ models in the infrared is also small.  What seemed like a big difference in NIR color-color space is actually a small difference in spectral space relative to $K$.

Variations in normalized extinction curve shape tend to be correlated across a wide spectral range because physical conditions that modify grains act efficiently across grain size distribution.  Extinction curve shape and the amount of extinction are not necessarily correlated.  While it is true that models with large amounts of visual extinction are steeper at all wavelengths than models with small extinction, a steep extinction curve and therefore large offset in color-color space do not necessarily indicate large extinction.  
It can be helpful to rewrite color excess as
\begin{equation}
    E(\lambda-{\rm ref})=A_{\rm ref} \left(\frac{A_\lambda}{A_{\rm ref}} -1\right),
\end{equation}
to emphasize that color excess measures both the amount of extinction ($A_{\rm ref}$) and the shape of the extinction curve ($A_\lambda/A_{\rm ref}$). The ratio of extinction to reddening, such as $R_V=A_V/E(B-V)$ in the blue-visual, is thus a measure of extinction curve shape only. 

\begin{figure*}[!tb] 
    \centerline{\includegraphics[angle=0,width=2\columnwidth]{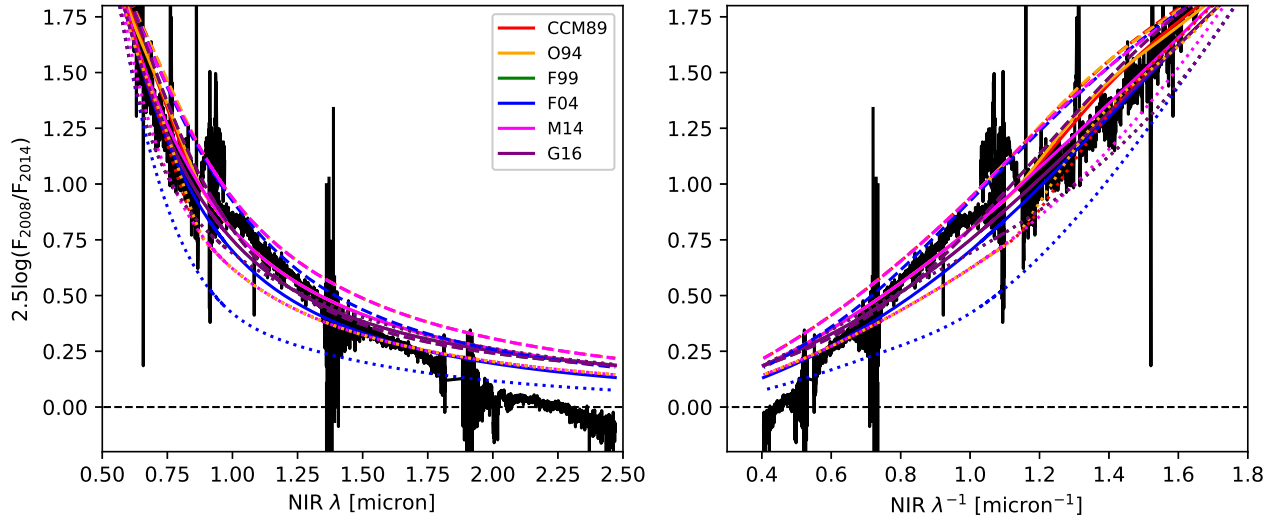}}
    \caption{Comparison of the NIR AA Tau extinction event to various extinction models {\bf $A_V=2$}.  Solid lines are models with $R_V=3$, consistent with the Milky Way average.  Models with $R_V =2$ typical of small grains are shown with dotted lines; $R_V=5$ typical of large grains are shown in with dashed lines. Note that none of these models adequately describe the apparent NIR extinction of AA Tau, primarily because of the steepness of the empirical curve and near-zero extinction at $K$.
    } 
    \label{fig:AATau_models}
\end{figure*}

One way to produce a steep color excess curve model is by invoking more extinction at all wavelengths by choosing a large $A_V$, in this case an overall scaling of $A_V=4$.  The other mechanism for making a steep extinction curve is invoking a different grain model, one that uses a different grain size distribution or optical properties, such that normalized ($A_\lambda/A_{\rm ref}$) is steep.  The commonly-used F99 extinction model, however, does not have any variety in NIR extinction curve shapes, that is, does not vary with $R_V$ in the near infrared{\footnote{This point has sometimes been misunderstood in the literature.  The tables in F99 give uniform spline anchors for $A_\lambda/E(B-V)$ in the near infrared, which would give values of $A_\lambda/A_V$ that depend strongly on $R_V$.  The text of the paper, however says that the spline anchors should be divided by $R_V$ first, which yields a $A_\lambda/A_V$ curve that is indeed independent of $R_V$.}}.  Therefore, invoking large extinction that scales the extinction curve at all wavelengths is the only mechanism for creating a steep NIR extinction curve with F99.

\subsection{Dust alone cannot explain AA Tau's 2008-2014 (apparent) differential extinction}

Could the problem be with our choice of extinction model? We use the F99 extinction model in this study because of its popularity as the source 
of the FMUNRED procedure in IDL, but there are other extinction laws (see, for 
example, a review of NIR extinction curve shapes in \citealt{Schlafly2016}).
In Figure \ref{fig:AATau_models}, we show $A_V=2$ curves for several other $R_V$-dependent models of Milky 
Way extinction: the well-known parameterization of CCM89 \citep{Cardelli1989}, the 
\citet{ODonnell1994} model that updates the optical portion of the CCM89 model, the
\citet{Fitzpatrick2004} update of F99 with a NIR $R_V$ dependence, the \citet{MaizApellaniz2014} model that smoothes  CCM89, and the 
two-parameter model of \citet{Gordon2016}. We use the \texttt{dust\_extinction}
affiliated package of Astropy to access these extinction curves.  None of these curves are good fits to AA Tau in the 
near infrared.  Most of the commonly used NIR extinction curves are forms of a power 
law for $A_\lambda$ or an offset power law for $E(\lambda-V)$
\citep[e.g.][]{MartinWhittet1990,Fitzpatrick2007}.
All power-law extinction shapes will be intrinsically unable to fit the nearly linear shape 
of the AA Tau extinction curve. Furthermore, because the power law shape approaches zero at 
long wavelengths, these models require some non-zero extinction at $K$.     
For AA Tau, apparent extinction in the $K$ band is nearly zero -- indeed, the extinction appears to become negative for wavelengths $\lambda > 2.25$ microns, a point we will return to shortly -- and 
therefore cannot be well fit by a power law.
We conclude that no reasonably expected model will fit AA Tau's NIR extinction in way that is consistent 
with what is observed in the visible.  

\begin{figure}[!htb] 
    \centerline{\includegraphics[angle=0,width=\columnwidth]{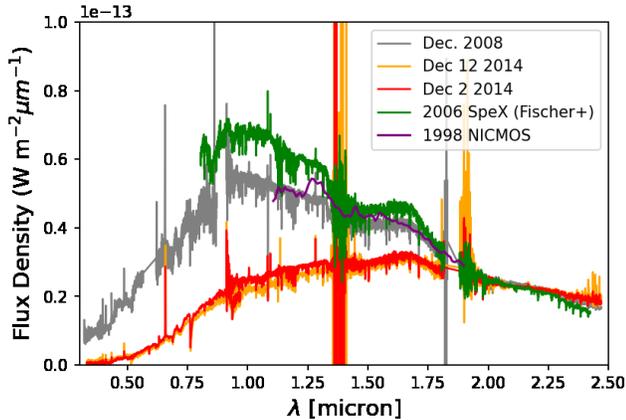}}
    \caption{Optical/NIR Spectra of AA Tau before and during the dimming event that commenced in late 2010.  NIR spectra were obtained in 2008 and 2014 with APO/Triplespec; contemporaneous optical spectra were obtained with Palomar/Doublespec (2008) and UH88/SNIFS (2014). Spectra obtained in 2006 by \citet{Fischer2011} with IRTF/SpeX and in 1998 by \citet{Carr1997} with HST NICMOS are shown for comparison. Note that all spectra converge to a similar flux density at $\lambda \sim 2.25$ microns, simultaneously supporting their independently performed flux calibrations, while also suggesting that AA Tau's $K$ band brightness remained essentially constant during the post-2011 dimming event, despite the dramatic changes in the blue portion of the spectrum.  Indeed, close inspection shows that AA Tau actually brightened between 2008 and 2014 at $\lambda > 2.25$ microns, and even more notably since 2006, implying the presence of additional infrared flux that is anti-correlated with the increase in extinction required to explain the optical behavior.} 
    \label{fig:AATau_fluxcal}
\end{figure}

Could grey (wavelength-independent) extinction be involved?  Increased grey extinction would be an additive factor in the apparent extinction $A_\lambda$ but would cancel out of the color excess $E(\lambda-V)$ or $E(\lambda-K)$.  Our finding that the shape of the AA Tau NIR extinction curve is unusual is based on the steepness of both the NIR extinction curve and the NIR color excess curve, and therefore cannot be explained by grey extinction.  In addition, grey extinction increases extinction at all passbands, including $K$.  Because the empirical extinction in the $K$ band is already so small, including grey extinction in the model would make the mismatch worse, not better.  The observed extinction at $K$ thus sets a low upper limit on the possible amount of grey extinction.

We can also eliminate error in the flux calibration as a possible cause of the anomalous shape of the NIR extinction curve. Figure  \ref{fig:AATau_fluxcal} shows additional archive spectra of AA Tau, all of which agree with our observation of near-zero extinction at $\lambda \sim 2.25$ microns.  Even if an error were made in the overall scaling of the spectra, that error would be present as a uniform additive quantity at all wavelengths of log-space  $A_\lambda$. As in the case of grey extinction, color excess is independent of an overall scaling factor because in calculating the difference between magnitude extinction in two bands, the additive term cancels out.  We see the same straight shape of the NIR portion of both the extinction and the color excess curve.  Correcting for some sort of systematic scaling error would shift the extinction curve up or down, but the color excess curves would be unchanged and thus still not well fit by any of the standard extinction models.

\subsection{Veiling explains the apparent anomaly in AA Tau's 2008-2014 differential extinction curve}

We conclude that AA Tau's spectral evolution cannot be explained by simple extinction and, motivated by the increase in flux at $\lambda \sim 2.5$ microns that is visible from 2006 onward in Fig. \ref{fig:AATau_fluxcal}, consider the possibility that the 2011 dimming event was accompanied by an increase in the NIR emission from the inner disk.  Absorption line veiling measurements allow us to investigate this hypothesis.  We start by writing the observed flux $F_\lambda$ as 
\begin{equation}
    F_\lambda = \left( F_{\lambda *} + E_\lambda \right) 10^{-0.4A_\lambda} \label{eq:veiled_flux}
\end{equation}
where $E_\lambda$ is the veiling emission from the inner disk and $A_\lambda$ is the extinction from material beyond the inner disk.  The EW is the ratio between integrated line intensity,
\begin{equation}
    I = \int_{\rm line} \left( \left( F_{\rm cont*} + E_\lambda \right) - \left( F_{\lambda *} + E_\lambda \right) \right) 10^{-0.4A_\lambda} d\lambda
\end{equation}
and the line continuum,
\begin{equation}
    F_{\rm cont} = (F_{\rm cont*} + E_\lambda)10^{-0.4A_\lambda}.
\end{equation}
If we assume that veiling emission $E_\lambda$ and the extinction $A_\lambda$ change slowly in the line, we substitute averages $E$ and $A$, respectively.  With this approximation, the equivalent width simplifies to
\begin{equation}
    EW = \frac{\int_{\rm line} (F_{\rm cont*}-F_{\lambda *})d\lambda}{F_{\rm cont*}+E},
\end{equation}
which is independent of extinction.  The ratio of EWs in two epochs with different veiling excesses, $E_1$ and $E_2$, is
\begin{equation}
    \frac{EW_1}{EW_2} = \frac{F_{\rm cont*}+E_2}{F_{\rm cont*}+E_1}, \label{eq:ewratio}
\end{equation}
independent of extinction in either epoch.  The ratio of line intensities in two epochs, on the other hand, does not depend on the excess emission in either epoch and can be simplified to
\begin{equation}
    \frac{I_1}{I_2} = 10^{-0.4(A_1-A_2)} = 10^{-0.4\Delta A},
    \label{eq:iratio}
\end{equation}
with different extinctions, $A_1$ and $A_2$.  This difference in extinction from one epoch to another, $\Delta A$ is the event extinction curve we seek in this study.

\begin{figure*}[!htb] 
    \centerline{\includegraphics[angle=0,width=2\columnwidth]{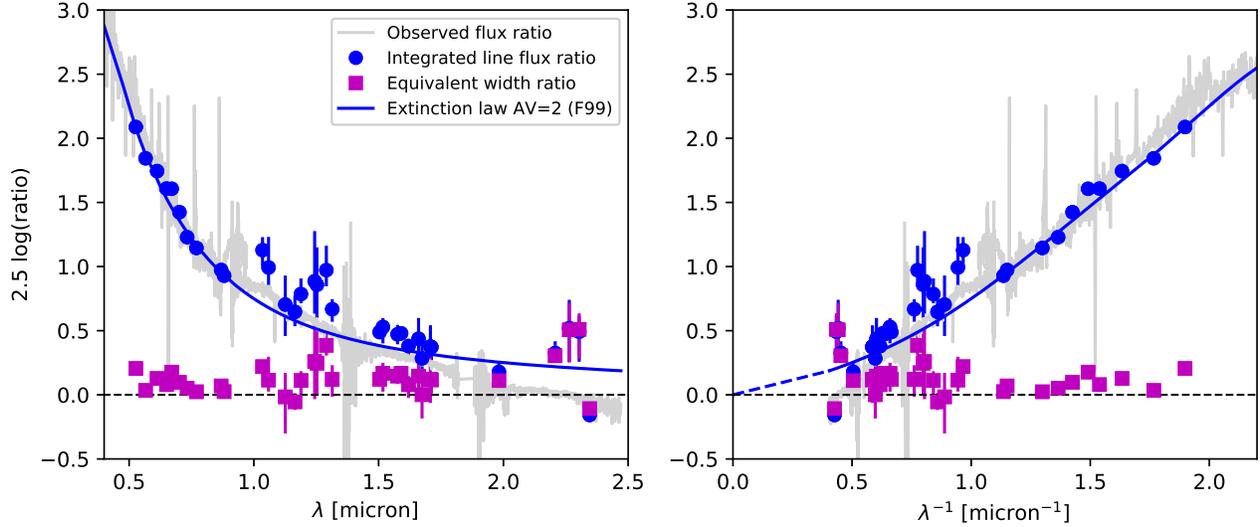}}
    \caption{AA Tau photospheric absorption line ratios from 2008 to 2014. Blue circles are integrated line flux ratios, $I_{2008}/I_{2014}$, and magenta squares are equivalent width ratios, $EW_{2008}/EW_{2014}$. Vertical lines indicate uncertainty estimated with repeat line measurements.  Observed flux ratio, $F_{2008}/F_{2014}$ is shown for comparison in grey.  Blue line is \citet{Fitzpatrick1999} extinction law for $A_V=2$, which agrees well with the line flux ratio.  Compare with Figure \ref{fig:AATau_ext} for similar format.} 
    \label{fig:veiling}
\end{figure*}

Figure \ref{fig:veiling} shows the ratios of line integrated intensities (blue circles) and EWs (magenta squares) between 2008 and 2014.  We plot the ratios in log space to illustrate how the extinction curve can be calculated from equation \ref{eq:iratio} above,
\begin{equation}
    \Delta A_{\lambda} = 2.5 \log\left(\frac{I_{2008}}{I_{2014}}\right). \label{eq:ext_Iratio}
\end{equation}
This method of using absorption lines to isolate extinction is similar to the method used by \citet{Fischer2011} to estimate extinction to AA Tau and other young stars relative to template spectra.

The blue model curve shown in Figure \ref{fig:veiling} is the FM99 extinction law for $A_V=2$.  For the most part, AA Tau's NIR line intensity ratios between these two snapshots in time are well represented by the F99 NIR extinction law, and therefore most power-law based extinction models as well.
We conclude that the flux ratio (shown in grey in Figure \ref{fig:veiling}), which until now was misinterpreted as an anomalous extinction curve, is actually the result of $A_V\simeq 2$, a common extinction curve shape, and veiling emission from the inner disk. 

\begin{figure}[!htb] 
     \centerline{\includegraphics[angle=0,width=\columnwidth]{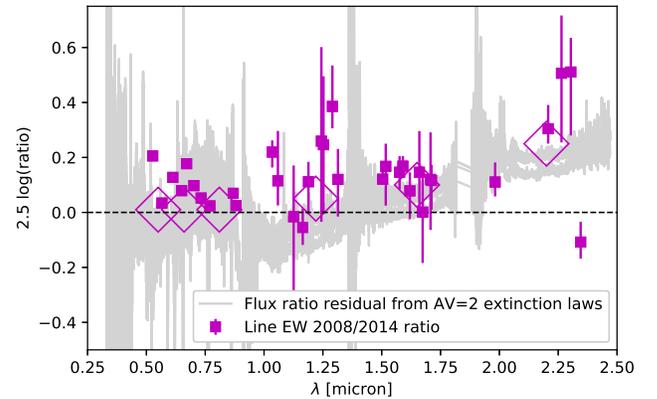}}
     \caption{AA Tau photospheric absorption line EW ratios.  As in Figure \ref{fig:veiling}, magenta squares are $EW_{2008}/EW_{2014}$.  Grey curves are differences between $A_V=2$ extinction curves plotted in Figure \ref{fig:AATau_models} and observed flux ratio in log space.  Note approximate agreement between EW ratio and residual from extinction curves.  Large magenta diamonds indicate representative values chosen for EW ratios in $VRI$ and $JHK$ passbands.}  
     \label{fig:veiling_justification}
\end{figure}

To test this hypothesis, Figure \ref{fig:veiling_justification}  shows the same EWs from Figure \ref{fig:veiling}  (magenta squares) superimposed on the difference between common extinction curves and the observed flux ratio.  From the definition of EW from equation \ref{eq:EWdef},
\begin{equation}
     2.5 \log\left(\frac{EW_{2008}}{EW_{2014}}\right) = 2.5 \log\left(\frac{I_{2008}}{I_{2014}}\right) - 2.5 \log\left(\frac{F_{2008}}{F_{2014}}\right)
\end{equation}
and substituting equation \ref{eq:ext_Iratio},
\begin{equation}
    2.5 \log\left(\frac{EW_{2008}}{EW_{2014}}\right) = \Delta A - 2.5 \log\left(\frac{F_{2008}}{F_{2014}}\right).
\end{equation}

We see that the EW ratios, which are the relative amount of veiling in the two epochs independent of extinction via equation \ref{eq:ewratio}, also approximately trace the difference between $A_V = 2$ extinction models and observed flux ratios.  Figure \ref{fig:veiling_justification} shows that this equivalency is generally true for AA Tau, confirming that veiling is the cause for the previously assumed ``anomalous'' extinction.  EW ratios are near unity in the optical and increase gradually through the near infrared, matching what we observed in the discrepancy between extinction models and observations in Figure     \ref{fig:AATau_spectra}, and indicating that veiling emission is negligible in the optical and becomes significant only in the near infrared.  Veiling flux increases through the near infrared, which is why the $A_V=2$ laws are a progressively worse fit to the flux ratios at longer wavelengths.

\subsection{Inferring increased NIR excess emission toward AA Tau during the dimming event}

It is natural to ask next what the veiling flux added between our 2008 and 2014 observations looks like.  In general, the question is not simple to answer because of how both the veiling in 2008 and the veiling in 2014 appear in the EW ratio, equation \ref{eq:ewratio}.  If we can assume that the veiling flux is small compared to the stellar flux, then the veiling ratio is approximately
\begin{equation}
    \frac{EW_{2008}}{EW_{2014}} = \frac{F_{\rm cont*}+E_{2014}}{F_{\rm cont*}+E_{2008}} \simeq 1 + \frac{E_{2014}-E_{2008}}{F_{\rm cont*}}.
\end{equation}
In this approximation, the ratio of EW is unity plus the increase in veiling as a fraction of the stellar photosphere.  When calculating relative to an unveiled spectrum, the veiling flux fraction is often called $r$. When calculating between two epochs, as we do here, the EW ratio indicates the relative change of the veiling flux fraction between the two observations.

A small-veiling approximation is appropriate for the 2008 and 2014 epochs of AA Tau at optical wavelengths.  \citet{Fischer2011} used unveiled template stars for veiling measurements of low- and high-resolution spectra of AA Tau obtained in 2006 and found a small veiling flux of $<10\%$ in the optical.  Similarly, we find line EW ratios near unity in the optical from 2008 to 2014, and thus can infer similarly low veiling flux in 2014; see Figure \ref{fig:opt_veiling}.

The picture is different in the infrared.  \citet{Fischer2011} found increasing NIR veiling flux of 20\% at 1 $\mu$m, 30\% at 2 $\mu$m, and 50\% for one line at 2.26 $\mu$m for AA Tau in 2006.
If veiling emission at the time of our observations in 2008 was similar to 2006, then we can calculate absolute veiling in 2014 directly from equation \ref{eq:ewratio} without the small-veiling approximation, where
\begin{equation}
    1+ \frac{E_{2014}}{F_{\rm cont*}}   = \left( 1 + \frac{E_{2008}}{F_{\rm cont*}} \right) \left( \frac{EW_{2008}}{EW_{2014}} \right). \label{eq:veilingemission}
\end{equation}
Our line EW ratios from 2008 to 2014 include substantial scatter, but we show as magenta diamonds in Figure \ref{fig:veiling_justification} representative values of 1.05, 1.10, and 1.25 for the 2008/2014 EW ratios we measure in the $J$, $H$, and $K$ bands, respectively.  Using equation 17 to calculate the $K$-band excess as a fraction of the photospheric flux with our representative value of $EW_{2008}/EW_{2014} =$ 1.25 and the \citet{Fischer2011} measurement of 30\% veiling flux at 2 microns in 2008  yields a 2014 veiling flux that is 60\% of the stellar flux at $K$. 
Using instead the \citet{Fischer2011} maximum of 50\% for the 2008 veiling emission at 2.26 $\mu$m, the result is $E_{2014}/F_{\rm cont*} \sim 1$.  In other words, half of the detected radiation in the $K$ band in 2014 may be veiling emission from the inner disk.

\begin{figure}[!tb] 
    \centerline{\includegraphics[angle=0,width=\columnwidth]{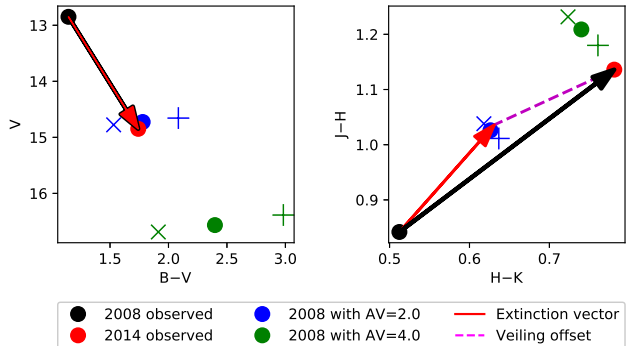}}
    \caption{Same as Figure \ref{fig:AATau_optnirplot}, but including effects of veiling.  The black vectors show AA Tau's total measured photometric change between 2008 and 2014, while the magenta vectors show the offsets due to the estimated veiling emission.  The red vector shows the difference between the total offset and the veiling offset.  The red vectors now agree well with the synthetic photometry of the 2008 spectrum reddened by a standard $A_V=2$ extinction law (blue circle).} 
    \label{fig:colormags_veiling}
\end{figure}

\begin{figure}[!tb] 
    \centerline{\includegraphics[angle=0,width=\columnwidth]{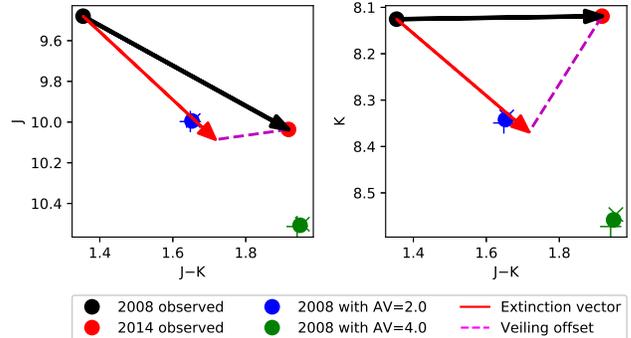}}
    \caption{NIR color-magnitude diagrams for AA Tau including effects of veiling.  As in Fig. \ref{fig:colormags_veiling}, black vectors show AA Tau's total measured photometric change between 2008 and 2014, while magenta vectors remove offsets due to estimated veiling emission.  Red vectors show the difference between the total offset and the veiling offset.  The red vectors agree with the synthetic photometry of the 2008 spectrum reddened by a standard $A_V=2$ extinction law (blue circle).}
    \label{fig:NIRcolormags}
\end{figure}

Finally, we return to the color-magnitude and color-color diagrams of Figure \ref{fig:AATau_optnirplot}.  We reinterpret the AA Tau's offset between 2008 and 2014 as due to the combined effects of increased emission from the inner disk and increased extinction.  Starting from equation \ref{eq:veiled_flux} and converting to log space, the magnitude offset becomes 
\begin{equation}
    \Delta m_\lambda = -2.5 \log\left( 1 + \frac{E_\lambda}{F_\lambda *}\right) + A_\lambda,
\end{equation}
a sum of two terms: a negative term for the veiling emission and a positive term for extinction.  The single vector offset in each panel of Figure \ref{fig:AATau_optnirplot} is actually a sum of two independent vectors, as we demonstrate in Figs. \ref{fig:colormags_veiling} and \ref{fig:NIRcolormags}.

From the analysis above, measurements of veiling flux relative to the stellar flux before the fade by \citet{Fischer2011} correspond to $\Delta J = -0.2$ and $\Delta K= -0.3$.  Our representative EW ratios in these bands (see magenta diamonds in Figure \ref{fig:veiling_justification}) correspond to $\Delta J = -0.05$ and $\Delta K= -0.25$ for the increased veiling flux in 2014 relative to 2008 before the fade.  In total, therefore, we estimate that in 2014, veiling flux contributed $\Delta J = -0.25$ and $\Delta K = -0.55$ relative to the stellar flux. 
Because there is no explicit line measurement in the $H$ band in \citet{Fischer2011} we interpolate between $J$ and $K$ and estimate that $\Delta H$ is  $-0.35$ relative to the stellar flux.  Figure \ref{fig:colormags_veiling} repeats Figure \ref{fig:AATau_optnirplot}, now with a magenta vector that removes the effect of veiling emission.  The red vector shows the difference between the total offset and the veiling offset, and is thus the empirical extinction vector for AA Tau's dimming event between the two epochs of our observations.  Keeping in mind the substantial uncertainties in our estimates, the empirical extinction vector appears to be fully consistent with the $A_V=2$ NIR models.  The extinction vectors in the optical color-magnitude diagram and the NIR color-color diagrams now agree.  We show the NIR color-magnitudes for $J$ and $K$ in Figure \ref{fig:NIRcolormags}. The dimming of AA Tau in the $J$ band from 2008 to 2014 was consistent with $A_V=2$ extinction laws, and the small increase in veiling emission at $J$ during that time does not change that conclusion.  Significant increased veiling at $K$, however, brings both $K$ color and magnitude closer to predictions based on $A_V =2$. \\

\section{Discussion, or, Rampant Speculation}

We now discuss the implications of AA Tau's standard extinction law and broad optical-mid-IR anti-correlation for the properties and location of its circumstellar material. 

\subsection{Dust grain size distribution \& dust-to-gas ratio}

Previous studies probing AA Tau's extinction law and/or inner disk composition have found conflicting results regarding the nature of its dust and gas content.  

Several studies analyze short-wavelength tracers captured in an HST/COS spectrum obtained in January 2011 (ie, immediately prior to, or just entering, the deep fade). From self-absorption in the Ly $\alpha$ profile,  \citet{McJunkin2014} measure an \ion{H}{1} column density that corresponds to $A_V=$0.34-0.61 for $R_V=$3.1-5.5, respectively: though this measurement does agree with the optically derived $A_V=$0.49 measurement by \citet{Kenyon1995}, they suggest that an enhanced dust-to-gas ratio may explain the systematic offset they detect between their HI derived extinction estimates and the preferentially higher $A_V$ values inferred from optical or NIR photometry.   
\citet{McJunkin2016} then re-analyze the HST/COS spectrum to model the excitation of H$_2$ emission lines via optical pumping from the Lyman $\alpha$ transition, inferring the UV extinction curve from the attenuation of the observed lines relative to the unobscured model.  From this analysis, \citet{McJunkin2016} infer a best fit extinction law with $R_V=2.6$ and a line-of-sight extinction of $A_V=$1.3 mag, but note the strong degeneracy between $R_V$ and $A_V$ over their restricted wavelength range.  Comparing their $H_2$ results to their prior HI analysis, \citet{McJunkin2016} note: 1) an overall systematic shift to higher $A_V$ values; 2) evidence of an enhanced dust-to-gas ratio; and 3) based on the smaller $R_V$ value they infer, suggestive evidence that AA Tau's sightline may be preferentially populated by small grains, and thus probing the upper levels of the (mid-fade) disk surface.  

The first intensive study of the dust properties associated with the newly obscuring material was conducted by \citet{Schneider2015}, examining the 2011 HST/COS spectrum along with a second HST/COS epoch from 2013 (obtained post-fade), and post-fade X-ray and NIR photometry and spectroscopy.  \citet{Schneider2015} note the distinct difference between the extinction $A_V=2$ implied in the $V-J$ color-magnitude diagram, and the $A_V=4.5$ implied by the source's motion in NIR color-color space, but follow \citet{Bouvier2013} in attributing the difference to the influence of scattering at optical wavelengths, and identifying the true additional line-of-sight extinction as $A_V=4.5$. of obscuring material. Comparing their X-ray derived N$_{HI}$ values to their optically (or NIR) inferred $A_V$ value, they also infer a dust-to-gas ratio close to the average interstellar value, which they interpret as evidence that the newly obscuring material likely lies in the outer part of the disk, where dust-to-gas ratios would be unlikely to depart from the standard interstellar value.

Our analysis focuses on AA Tau's SED evolution between the V and K band filters; in that regime, we find that changes in veiling (primarily in the H \& K bands) and an additional A$_V$=2 of extinction that follows a standard extinction law are sufficient to reproduce the changes in AA Tau's SED.  Our data therefore does not require scattering to reproduce the SED for $\lambda > 5000$ Angstroms, but neither does it rule out or provide strong constraints on the contribution of scattering at shorter wavelengths ($\lambda < 4000$ Ang.), where scattering is most effective and where the HST/COS observations of H2 lines analyzed by \citet{Schneider2015} are more sensitive to contributions by scattering. 

With no need to invoke a non-standard extinction law to explain AA Tau's photometric and spectroscopic evolution, we also find no evidence that the grain size distribution of the newly obscuring matter deviates from that assumed in the standard interstellar ($R_V= 3.1$) extinction law.  Relative to the modestly lower $R_V=2.6$ extinction law inferred by \citet{McJunkin2016}, however, the new material may still represent a shift in the dust grain size distribution to somewhat larger grains. Near uniformity of NIR extinction models and relative insensitivity to $R_V$ in the NIR makes it difficult to draw such conclusions.  If true, this picture could be consistent with the 2011 sightline probing the disk's uppermost layer, where the largest grains may have settled out.  If the disk's scale height increases to cause the 2011 fade, as we advocate below, this newly obscuring material would be drawn from further down in the disk, and thus less depleted in large grains and more consistent with the standard interstellar extinction law.  The degeneracy between $A_V$ and $R_V$ in the \citet{McJunkin2016} analysis makes it difficult to conclusively prove, however, that the new material is truly enhanced in larger grains relative to the original sightline: studies with archival spectra or SED information that could break the $A_V$ and $R_V$  degeneracy in the $H_2$ analysis would be welcome. Similarly, a differential study of the $H_2$ emission lines in the 2011 and 2013 HST/COS spectra (or new observations planned for Cycle 28, PI Schneider) may provide useful and self-consistent constraints on the amount and wavelength dependence of the extinction associated with the newly obscuring material.  

Our results do not directly probe the dust-to-gas ratio, but our results suggest that AA Tau may also possess more pedestrian dust-to-gas ratios than previously inferred from attributing to extinction the reddening that results from intrinsic mid-IR brightening, and thus artificially inflating the disk's dust content. \\ \\

\subsection{Location of the extinction}

Two pieces of evidence suggest that the dust responsible for the additional $A_V=2$ of extinction entering the line of sight to AA Tau in $\sim$2011 is located in an azimuthally asymmetric warp in the disk at a few AU.  First, \citet{Bouvier2013} assumed that the disk itself is not changing, and that a static warp in the disk rotated into our view.  Keplerian rotation and the long-lasting bright period places the warp at a distance of $>7.7$ AU.  Second, the decrease in flux of far-ultraviolet H$_2$ lines at high velocities also suggests that the excess absorber is located at $\sim2$ AU \citep{Schneider2015}.

However, if the absorbing material is located at $\sim2$-8 AU, then we should expect that extinction should affect all emission from AA Tau and its inner disk; AA Tau would be fainter at all wavelengths into the thermal infrared during this fade.  In contrast to this expectation, we detect near-IR and mid-IR brightening during this fade, which is more easily explained by absorbing material in the inner disk.

\begin{figure}[!tb] 
    \centerline{\includegraphics[angle=0,width=\columnwidth]{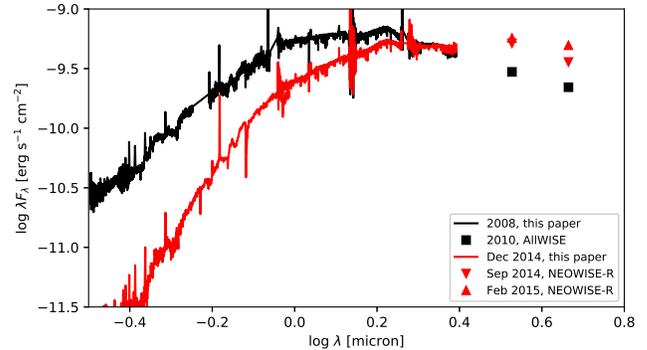}}
    \caption{Spectral energy distribution of AA Tau. AllWISE and NEOWISE-R data near the epochs of the spectra are from Table 1. } 
    \label{fig:SED}
\end{figure}

AA Tau's infrared brightening is detected in several ways in our analysis. Figures \ref{fig:AATau_models} and \ref{fig:AATau_fluxcal} show that AA Tau was actually brighter in the $K$-band when it was much fainter at optical wavelengths.   Figures \ref{fig:veiling} and \ref{fig:veiling_justification} show that the $K$-band veiling increased, indicating that the brightening is likely driven by flux produced by the disk and not the star.
To confirm the presence of this NIR excess, and its increase in strength during the period of AA Tau's optical dimming, we supplement our 2008 and 2014 optical/NIR spectra with mid-IR fluxes calculated from Spitzer and WISE photometry.  The resultant 2008 and 2014 SEDs are shown in Figure \ref{fig:SED}, and unambiguously demonstrate the presence of a significant mid-IR excess whose  flux increased in brightness in absolute units from before the optical fade to after the optical fade. Similar mid-IR excesses, and their anti-correlation with apparent optical extinction, have been previously reported for other dipper/UX Or-like pre-main sequence stars \citep[e.g.,][]{Grinin2009, Shenavrin2015, Koutoulaki2019}, so are not unique to the AA Tau system.

\begin{figure}[!tb] 
    \centerline{\includegraphics[angle=0,width=\columnwidth]{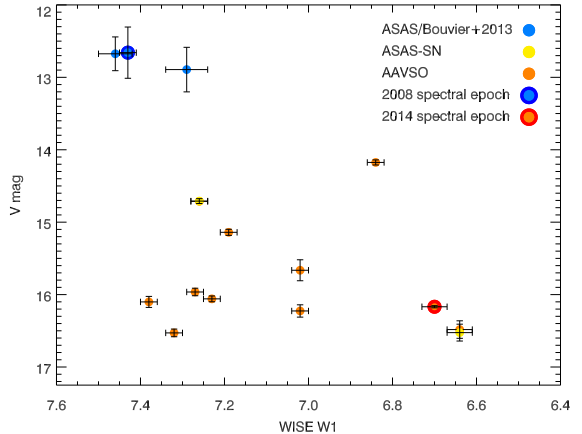}}
    \caption{WISE W1 vs. $V$ (measured from AAVSO, ASAS, or \citet{Bouvier2013}) for AA Tau over the period 2005 - 2019 (2005 epoch utilizes Spitzer IRAC [3.4] as a pseudo W1 measurement). } 
    \label{fig:WISEvsV}
\end{figure}

The brightening that is barely detected at the long-wavelength end of the $K$ band is more prominent at longer wavelengths: as Figures \ref{fig:SED} and \ref{fig:WISEvsV} show, AA Tau's 3--5 $\mu$m emission, as measured with {\it WISE} and {\it Spitzer}, are faintest when AA Tau is optically bright and are brighter (albeit with substantial scatter in the WISE light curve) when AA Tau is optically faint.  

AA Tau's non-photospheric $2-5~\mu$m emission is produced in the innermost disk, where the dust is warmest \citep[e.g.][]{dalessio98}. For an inner disk that is observed nearly edge-on, most of the detected near-IR and mid-IR disk emission would be emitted from the far side of the disk, with the warm inner disk wall dominating at the shortest wavelengths. The location of the inner disk wall,  where dust begins to sublimate at T $\gtrsim$ 1400 K \citet{Kobayashi2011}, can be estimated by following \citet{Whitney2004}: 
\begin{equation}
    R_{sub} = R_* \bigg( \frac{T_{sub}}{T_{eff}} \bigg)^{-2.1}
\end{equation}
and adopting $R_*$ = 1.8 $R_{\odot}$ and $T_{eff}$ = 4000 K for AA Tau \citep{Esau2014}.  This predicts AA Tau's inner disk wall lies at $\sim$7 $R_*$ = $\sim$12.65 R$_{\odot}$ = $\sim$0.06 AU, consistent with locations inferred by \citet{Esau2014}, on the basis of detailed radiative transfer modelling, and a similar calculation by \citet{Bodman2017}, who find AA Tau's inner disk extends slightly within the disk's co-rotation radius.  Any brightening at mid-IR wavelengths likely indicates an increase in the visible surface area of the warm inner disk; in other words, the inner disk wall has increased in height.  This taller inner disk would also, on its near side, then naturally lead to more obscuration of the central star. 
 
In this interpretation, the scatter in the mid-IR photometry would indicate temporal changes in the inner disk scale height.  The scatter in the $V$-band lightcurve also demonstrates that the optical depth of the absorber changes with time.  In this geometry we should expect a stronger anti-correlation between the $V$-band and 3--5 $\mu$m emission, azimuthal asymmetries in the inner disk were detected before the fade \citep[e.g.][]{Bouvier2007} and may persist.  The 3--5 $\mu$m emission should be dominated by the back side of the disk, while the optical extinction is caused by the front side of the disk.  Moreover, the optical depth and geometric height of the inner disk may not exactly correlate. 
 
Such anti-correlated mid-IR and optical lightcurves have been previous detected and modeled as changes in the inner disk \citep{mcginnis2015}.  This seesaw effect has also been seen with a pivot point at longer wavelengths but with a similar physical attribution of a change in height of the inner disk \citep{Espaillat2011,flaherty12, Bryan2019}.  The proximate cause of such a change is uncertain.  We speculate that a small change in the surface density of the inner disk would lead to a small increase in disk height. The increased surface density could also correspond with a small increase in accretion rate, causing the truncation radius to contract \citep[see review by][]{hartmann2016}, in which case the disk would cover a larger fraction of the star.  Since the inner disk of AA Tau was already viewed with a high enough inclination to occasionally obscure the star \citep[e.g.][]{Bouvier2007}, it would not be surprising if small changes in the geometry of the inner disk led to a persistent obscuration.

The inner disk location of the obscuration would also be more consistent with the detection of surface disk accretion by \citet{zhang2015}.  If the redshifted CO absorption traces a blob moving at a constant speed of 6 km/s, over 10 years the blob would have moved $\sim 13$ AU and may have rotated out of our line of sight to the star.  If the obscuration is in the inner disk, then the redshifted absorption should instead be interpreted as a steady accretion flow associated with the inner wall at the disk truncation radius; additional modelling would be useful for exploring the range of accretion rates and blob volume densities that can self-consistently reproduce the extinction and molecular line measurements.

An alternative interpretation for the near-IR variability would need to invoke an increase in the surface area of the warm inner disk.  The surface area of the warm inner disk could expand if either viscous heating or irradiation of the disk by the central star increased significantly.  Both of these possibilities would require a sustained and large increase in the accretion rate, but veiling measurements indicate that any difference between the accretion rate before and during the fade are modest \citep[see also][]{Bouvier2007,Schneider2015}.

\section{Conclusions}

AA Tau dimmed by $A_V= 2.0$ between December 2008 and December 2014.  We use contemporaneous optical and NIR spectra to calculate an empirical differential extinction curve between these two dates, bracketing the dimming event.  We conclude the following: 

1. It is not possible with regular, commonly-used extinction curve families to reproduce both the optical and the infrared spectral regimes of this dimming event simultaneously if extinction is considered to be the only cause of differences between the two epochs.

2. Measuring the equivalent width and integrated strength of absorption lines allows \textit{independent measurements} of the change in veiling flux and the change in extinction, respectively.  We measure increases in veiling flux into the NIR that are largest at $K$, where the veiling flux is 60\% or more of the stellar flux in 2104, double what \citet{Fischer2011} measured for AA Tau before the fade.

3.  When corrected for increased NIR veiling emission, the empirical extinction curve of AA Tau from before to after the 2011 dimming event can be described by regular, commonly-used ($R_V\sim3$) extinction curves and an increase of $A_V=2$.  The pedestrian nature of the extinction law associated with this fade also suggests that the material newly added along the line of sight to AA Tau has a dust grain distribution that is broadly consistent with standard interstellar dust, is not strongly affected by either settling or radiative processing in the inner disk, and does not point to highly enhanced dust-to-gas ratios in the inner disk. 

4. Our NIR spectra, along with mid-IR photometry from WISE and Spitzer, indicate that AA Tau's infrared brightness increased as its optical brightness declined in its extended fade.  This observational signature, detected previously in other UX Or like variables, is consistent with predictions of scale height changes in the inner disk.  The timescale of AA Tau's current fade implies that such scale height changes can persist for decades. 

5. Previous work using NIR color-color diagrams of AA Tau should be interpreted as detecting an unusually steep apparent differential extinction curve in the NIR, not a factor of two more extinction, as has been reported. We show the importance of obtaining spectral data for studying dimming events and the importance of considering veiling.  We recommend caution in inferring extinction of young stars from color-color diagrams alone. \\

\acknowledgements

The authors thank Christian Schneider, Jerome Bouvier, and the referee, Vladamir Grinin, for helpful reviews of the submitted draft that improved the content and presentation of this analysis.  K.R.C thanks Will Fischer for insightful discussions on methods to measure, disentangle, and interpret veiling and extinction in T Tauri star spectra, and J. Rodriguez for assistance in interpreting KELT light curves.  G.J.H. thanks Carlos Contreras Pena and Wooseok Park for discussions of neoWISE light curves.   K.R.C acknowledges support provided by the NSF through grant AST-1449476, and thanks the Time Allocation Committee in the University of Washington's Astronomy Department for supporting the acquisition of the near-infrared spectra presented here. G.J.H. is supported by general grant 11773002 awarded by the National Science Foundation of China.
This research made use of Astropy, a community-developed core Python package for Astronomy \citep{astropy2018}.

\facilities{AAVSO, ARC, Hale}  

\software{Astropy}

\bibliography{AATau}

\begin{splitdeluxetable*}{ccccccccccccBccccccccccc}
\rotate
\tablewidth{0pt}
\tabletypesize{\tiny}
\tablecaption{Optical Veiling Measurements\label{tab:Opt_Veil}}
\tablehead{
  \colhead{} &
  \colhead{} &
  \colhead{} &
  \colhead{5270} &
  \colhead{5270} &
  \colhead{5660} &
  \colhead{5660} &
  \colhead{Ca 6122} &
  \colhead{Ca 6122} &
  \colhead{6500} &
  \colhead{6500} &
  \colhead{Li 6708} &
  \colhead{Li 6708} &
  \colhead{K 7016} & 
  \colhead{K 7016} &
  \colhead{K 7326} &
  \colhead{K 7326} &
  \colhead{K 7700} &
  \colhead{K 7700} &
  \colhead{Fe 8689} &
  \colhead{Fe 8689} &
  \colhead{Mg 8807} &
  \colhead{Mg 8807} \\
  \colhead{MJD} &
  \colhead{$i$ flux} &
  \colhead{$i$ Mag} &
  \colhead{Flux} &
  \colhead{EqW} &
  \colhead{Flux} &
  \colhead{EqW} &
  \colhead{Flux} &
  \colhead{EqW} &
  \colhead{Flux} &
  \colhead{EqW} &
  \colhead{Flux} &
  \colhead{EqW} &
  \colhead{Flux} & 
  \colhead{EqW} &
  \colhead{Flux} &
  \colhead{EqW} &
  \colhead{Flux} &
  \colhead{EqW} &
  \colhead{Flux} &
  \colhead{EqW} &
  \colhead{Flux} &
  \colhead{EqW} }

\startdata
54806.438 & 4.90 & -1.73 & 1.096E-02 & 4.636E-01 & 2.551E-03 & 9.462E-02 & 2.627E-03 & 8.833E-02 & 4.475E-03 & 1.397E-01 & 1.731E-03 & 5.754E-02 & 9.092E-04 & 2.553E-02 & 1.153E-03 & 3.076E-02 & 2.240E-03 & 5.565E-02 & 2.376E-03 & 4.777E-02 & 2.506E-03 & 4.688E-02 \\
54806.453 & 5.18 & -1.79 & 1.173E-02 & 4.650E-01 & 2.774E-03 & 9.634E-02 & 2.785E-03 & 8.806E-02 & 4.768E-03 & 1.402E-01 & 1.850E-03 & 5.758E-02 & 9.542E-04 & 2.532E-02 & 1.223E-03 & 3.079E-02 & 2.399E-03 & 5.618E-02 & 2.587E-03 & 4.906E-02 & 2.709E-03 & 4.774E-02 \\
54806.465 & 5.08 & -1.76 & 1.159E-02 & 4.697E-01 & 2.642E-03 & 9.409E-02 & 2.714E-03 & 8.797E-02 & 4.570E-03 & 1.382E-01 & 1.820E-03 & 5.850E-02 & 9.192E-04 & 2.498E-02 & 1.204E-03 & 3.100E-02 & 2.370E-03 & 5.633E-02 & 2.467E-03 & 4.791E-02 & 2.672E-03 & 4.793E-02 \\
54806.480 & 5.03 & -1.75 & 1.173E-02 & 4.755E-01 & 2.531E-03 & 9.133E-02 & 2.682E-03 & 8.719E-02 & 4.635E-03 & 1.405E-01 & 1.790E-03 & 5.789E-02 & 9.432E-04 & 2.567E-02 & 1.189E-03 & 3.083E-02 & 2.284E-03 & 5.537E-02 & 2.504E-03 & 4.885E-02 & 2.584E-03 & 4.713E-02 \\
54807.387 & 5.80 & -1.91 & 1.355E-02 & 4.630E-01 & 3.052E-03 & 9.209E-02 & 3.236E-03 & 8.840E-02 & 5.471E-03 & 1.404E-01 & 2.094E-03 & 5.715E-02 & 1.156E-03 & 2.676E-02 & 1.397E-03 & 3.115E-02 & 2.723E-03 & 5.603E-02 & 2.915E-03 & 4.937E-02 & 3.035E-03 & 4.820E-02 \\
54807.402 & 5.82 & -1.91 & 1.387E-02 & 4.721E-01 & 3.255E-03 & 9.739E-02 & 3.317E-03 & 8.988E-02 & 5.545E-03 & 1.417E-01 & 2.109E-03 & 5.746E-02 & 1.143E-03 & 2.657E-02 & 1.416E-03 & 3.137E-02 & 2.698E-03 & 5.552E-02 & 2.877E-03 & 4.855E-02 & 3.120E-03 & 4.906E-02 \\
54807.418 & 5.89 & -1.93 & 1.388E-02 & 4.647E-01 & 3.034E-03 & 8.984E-02 & 3.289E-03 & 8.852E-02 & 5.557E-03 & 1.404E-01 & 2.128E-03 & 5.716E-02 & 1.144E-03 & 2.631E-02 & 1.411E-03 & 3.103E-02 & 2.671E-03 & 5.435E-02 & 2.922E-03 & 4.892E-02 & 3.171E-03 & 4.937E-02 \\
54807.430 & 5.89 & -1.93 & 1.407E-02 & 4.701E-01 & 3.236E-03 & 9.536E-02 & 3.270E-03 & 8.778E-02 & 5.514E-03 & 1.389E-01 & 2.129E-03 & 5.708E-02 & 1.215E-03 & 2.768E-02 & 1.394E-03 & 3.063E-02 & 2.745E-03 & 5.567E-02 & 2.934E-03 & 4.893E-02 & 3.090E-03 & 4.809E-02 \\
54808.406 & 4.09 & -1.53 & 8.372E-03 & 4.620E-01 & 1.958E-03 & 9.359E-02 & 2.076E-03 & 8.798E-02 & 3.628E-03 & 1.407E-01 & 1.394E-03 & 5.745E-02 & 7.306E-04 & 2.526E-02 & 9.363E-04 & 3.049E-02 & 1.897E-03 & 5.667E-02 & 1.932E-03 & 4.595E-02 & 2.163E-03 & 4.725E-02 \\
\nodata \\
57036.449 & 0.20 & 1.77 & 1.544E-04 & 3.994E-01 & 4.210E-05 & 8.636E-02 & 5.628E-05 & 8.453E-02 & 1.139E-04 & 1.393E-01 & 4.461E-05 & 5.391E-02 & 2.612E-05 & 2.361E-02 & 3.869E-05 & 2.881E-02 & 8.555E-05 & 5.825E-02 & 1.074E-04 & 4.373E-02 & 1.203E-04 & 4.427E-02 \\
57036.461 & 0.20 & 1.77 & 1.415E-04 & 3.737E-01 & 4.182E-05 & 8.606E-02 & 5.712E-05 & 8.472E-02 & 9.744E-05 & 1.236E-01 & 4.421E-05 & 5.315E-02 & 2.368E-05 & 2.151E-02 & 4.099E-05 & 3.003E-02 & 8.355E-05 & 5.666E-02 & 1.128E-04 & 4.563E-02 & 1.201E-04 & 4.448E-02 \\
57037.191 & 0.23 & 1.59 & 1.935E-04 & 3.244E-01 & 5.209E-05 & 8.099E-02 & 5.665E-05 & 6.695E-02 & 1.100E-04 & 1.028E-01 & 3.884E-05 & 3.321E-02 & 2.832E-05 & 1.928E-02 & 5.010E-05 & 2.903E-02 & 8.856E-05 & 4.731E-02 & 1.103E-04 & 3.967E-02 & 1.212E-04 & 4.037E-02 \\
57037.203 & 0.20 & 1.75 & 1.529E-04 & 3.897E-01 & 4.246E-05 & 8.446E-02 & 5.588E-05 & 8.168E-02 & 9.832E-05 & 1.193E-01 & 4.348E-05 & 5.184E-02 & 2.002E-05 & 1.788E-02 & 4.302E-05 & 3.132E-02 & 8.882E-05 & 5.841E-02 & 1.077E-04 & 4.311E-02 & 1.157E-04 & 4.272E-02 \\
57037.215 & 0.20 & 1.77 & 1.519E-04 & 3.978E-01 & 4.043E-05 & 8.181E-02 & 5.614E-05 & 8.374E-02 & 9.793E-05 & 1.221E-01 & 4.283E-05 & 5.170E-02 & 2.154E-05 & 1.958E-02 & 3.917E-05 & 2.935E-02 & 9.252E-05 & 6.123E-02 & 1.088E-04 & 4.457E-02 & 1.185E-04 & 4.444E-02 \\
57037.227 & 0.20 & 1.75 & 1.464E-04 & 3.841E-01 & 4.588E-05 & 8.964E-02 & 5.539E-05 & 8.129E-02 & 1.011E-04 & 1.252E-01 & 4.492E-05 & 5.224E-02 & 2.762E-05 & 2.458E-02 & 4.155E-05 & 3.051E-02 & 9.506E-05 & 6.248E-02 & 1.031E-04 & 4.162E-02 & 1.214E-04 & 4.481E-02 \\
\enddata
\end{splitdeluxetable*}

\end{document}